\documentclass{getwriting}
\usepackage{graphicx} 
\usepackage{amsmath}
\usepackage[numbers,sort&compress]{natbib}

\title{Organ-scale pulsatile blood flow reconstruction in the avian yolk sac embryo}
\author[1]{Sidney Holden \thanks{These authors contributed equally to this work.}\thanks{sholden@flatironinstitute.org}}
\author[1]{Alexander Mayer \protect\footnotemark[1]\thanks{amayer@flatironinstitute.org}}
\author[2]{Siavash Ghaffari}
\author[3,4]{Elizabeth A.V. Jones}
\author[1,5]{Eleni Katifori}

\affil[3]{\footnotesize Department of Cardiovascular Sciences, KU Leuven, Leuven, Belgium}
\affil[4]{\footnotesize Department of Cardiology, CARIM School for Cardiovascular Diseases, Maastricht University, Maastricht, Netherlands}
\affil[2]{\footnotesize Department of Chemical Engineering, McGill University, Montreal, QC, Canada}
\affil[5]{\footnotesize Department of Physics and Astronomy, University of Pennsylvania, Philadelphia, PA, USA}
\affil[1]{\footnotesize Center for Computational Biology, Flatiron Institute, New York, NY, USA}

\begin{document}
\maketitle
\begin{abstract}
\noindent The hemodynamic forces on endothelial cells in developing vascular networks vary over the cardiac cycle.
Resolving these variations throughout the yolk sac requires imaging individual regions of the vasculature at different times.
We reconstructed mean and pulsatile blood flow in a 21-somite quail embryo from 53 overlapping high-speed microscopy recordings acquired over 3.7\,h.
We reconciled mean-flow measurements using conservation of flux at vascular junctions and synchronized cardiac phases using vessels shared between recordings.
Of the 7,009 vessels with flow measurements in the computational domain, we reconstructed pulsatile flow in 5,214.
Relative pulsatility decreased along an arterial-side path but increased along a venous-return path, where downstream flow oscillations preceded those upstream.
Across the plexus, phase lags relative to the vitelline veins increased toward the vascular margin.
A distributed resistance--compliance model with uniform vessel distensibility reproduced many of these spatial variations.
Our results suggest that vascular geometry and compliance, together with pulsatile forcing from both arterial and venous boundaries, determine the local flow waveforms within the developing plexus.
\end{abstract}

\section*{Introduction}

The hemodynamic forces behind vascular remodeling depend on vascular structure across scales, from the caliber of individual vessels to the organization of the network as a whole.
In vertebrate embryos, endothelial cells respond to hemodynamic forces as primitive vascular plexuses remodel into hierarchical networks \citep{le_noble_flow_2004,lucitti_vascular_2007,weijts_blood_2018,culver_effects_2010}.
With changes in vessel caliber and connectivity redistributing blood flow, remodeling in one part of the network changes the forces experienced by endothelial cells elsewhere.

The forces acting on the endothelium also change throughout the cardiac cycle.
These temporal variations matter independently of mean shear stress: changes in shear waveform and frequency can alter endothelial gene expression even when mean shear is held constant \citep{blackman_new_2002,himburg_frequency-dependent_2007,feaver_human_2013}.
Increasing the rate of rise of shear stress also accelerates endothelial alignment and elongation in vitro \citep{hsiai_endothelial_2002}.
In the developing chick yolk sac, this feature of the flow waveform distinguishes arteries from veins.
Arterial-type pulsatile shear also enhances arterial marker expression relative to constant shear \citep{buschmann_pulsatile_2010}.
Mean flow alone therefore provides an incomplete description of the hemodynamic signals experienced by endothelial cells throughout a developing vascular network.

Pulsatile flow in an individual vessel depends on both the heartbeat and the mechanical response of the vascular network.
Compliant vessels expand and recoil with each heartbeat, storing and releasing blood and changing the amplitude, timing, and shape of flow oscillations.
Experiments and modeling in zebrafish embryos have shown that arterial elasticity can alter the pulsatile flow generated by the heart \citep{anton_pulse_2013}.
In chick embryos, pulse propagation changes with developmental stage, and arterial compliance and impedance change in response to altered hemodynamic loading \citep{yoshigi1997developmental,lucitti2005arterial}.
Microvessel distensibility also changes as vessels mature in the chick chorioallantoic membrane \citep{durussel2005evolution}.
How these vessel properties combine across a developing plexus to shape pulsatile flow remains unclear.

Previous work characterizing blood flow across microvascular networks has focused primarily on mean flow.
In the chick yolk sac, micro-particle image velocimetry has measured time-averaged flow at different developmental stages, showing substantial variation among vessels of similar caliber \citep{kloosterman_quantification_2014}.
Network models have examined how vessel geometry and connectivity contribute to these flow distributions \citep{pries_structure_1995,maibier_structure_2016,smith_microvascular_2016}.
Optimization and probabilistic inference methods estimate flow when hemodynamic measurements are incomplete or boundary conditions are undetermined \citep{fry2012estimation,rasmussen2017model,rasmussen2025blood}.

Pulsatile flow presents an additional challenge, as measurements must resolve individual cardiac cycles.
In the chick vitelline network, phase-resolved velocity measurements have been used to estimate wall shear stress \citep{poelma2008vivo}.
In quail embryos, \citet{ghaffari_simultaneous_2015} combined time-resolved imaging with computational fluid dynamics to study local hemodynamics during vascular remodeling.
Cardiac oscillations have also been measured in the mouse cerebral microcirculation using two-photon microscopy \citep{santisakultarm2012vivo} and, more recently, laser speckle imaging \citep{jensen2025cardiac}.

In our experiments, the field of view needed to resolve flow in individual vessels covers only a small portion of the yolk sac.
Imaging the entire vasculature therefore takes several hours, over which the heart rate can change considerably.
Although vessels within a single recording share a common time reference, those in separate recordings do not.

Here, we reconstruct mean and pulsatile blood flow across the quail yolk sac at the 21-somite stage.
We use conservation of mean flow at vascular junctions and repeated measurements of vessels in overlapping recordings to reconstruct the flow field over a cardiac cycle.
We then investigate how well vessel geometry predicts the distribution of mean flow, and whether the distensibility of the vasculature can explain the changes in pulse amplitude and phase observed throughout the network.
\section*{Results}

\subsection*{Reconstruction of the vascular network from sequential microscopy}

\begin{figure}[t]
    \centering
    \includegraphics[width=\linewidth]{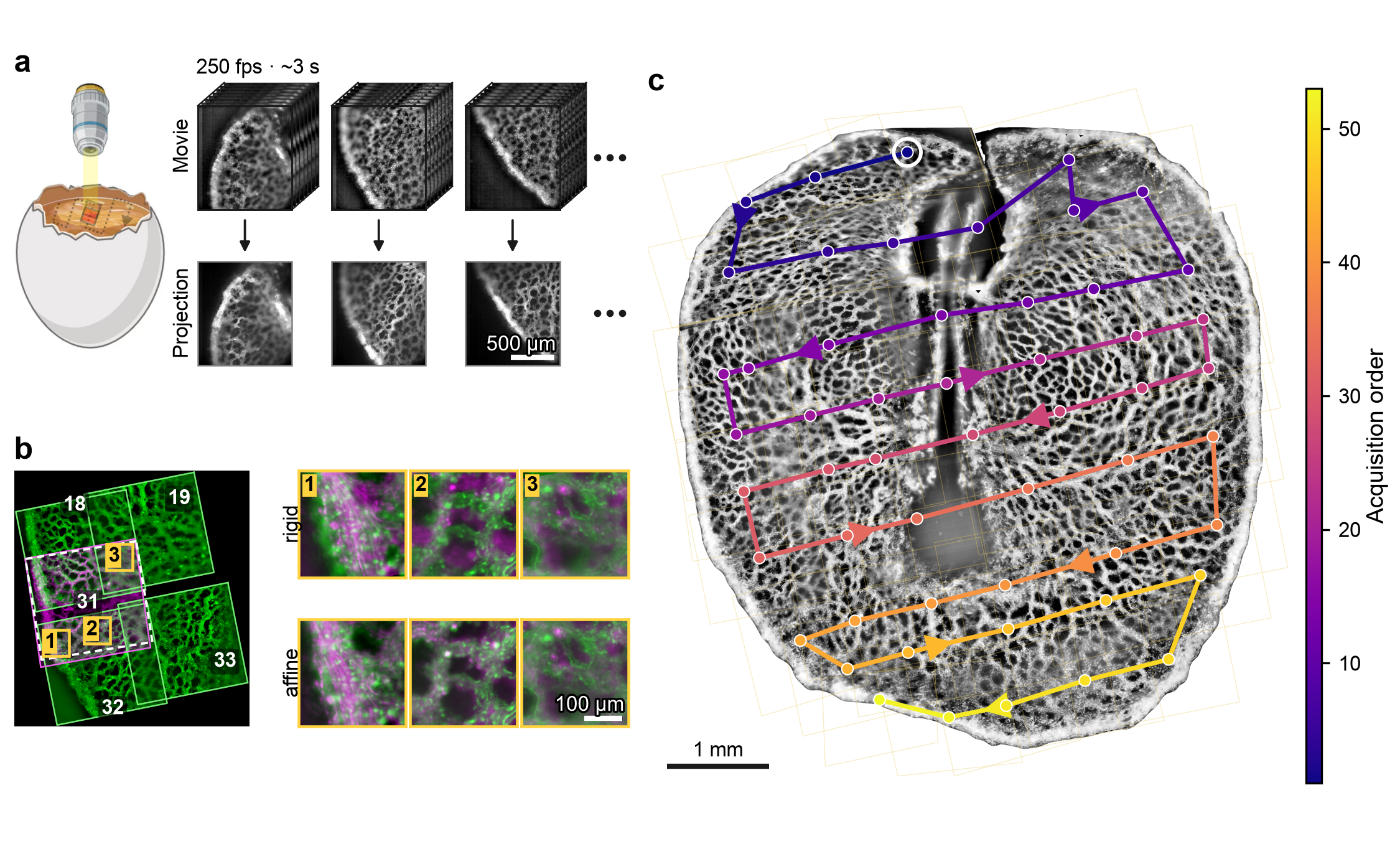}
    \caption{\textbf{Image acquisition and reconstruction of the vascular mosaic.}
    \textbf{a,} Schematic of the imaging setup. Individual regions were recorded at 250 frames\,s$^{-1}$ for approximately 3\,s, and each recording was reduced to a maximum-intensity projection.
    \textbf{b,} Comparison of rigid and affine registration between tile 31 (magenta) and neighboring tiles (green). Numbered boxes indicate three overlapping regions shown at higher magnification.
    \textbf{c,} Assembled vascular mosaic showing the positions and acquisition order of all 53 tiles.}
    \label{fig:acquisition}
\end{figure}

We analyzed 53 overlapping high-speed recordings of the yolk-sac vasculature of a 21-somite quail embryo (Fig.~\ref{fig:acquisition}a). Each recording captured several cardiac cycles at 250 frames\,s$^{-1}$. Acquisition of the full mosaic took approximately 3.7\,h, during which the vasculature continued to grow and remodel.

Although neighboring tiles overlapped spatially, they were not necessarily acquired close together in time. Changes in vascular geometry between recordings made rigid registration insufficient in some regions. We therefore applied affine transformations to 31 of the 53 tiles (Fig.~\ref{fig:acquisition}b). The assembled mosaic and the order in which the tiles were acquired are shown in Fig.~\ref{fig:acquisition}c.

\begin{figure}[!htbp]
    \centering
    \includegraphics[width=\linewidth]{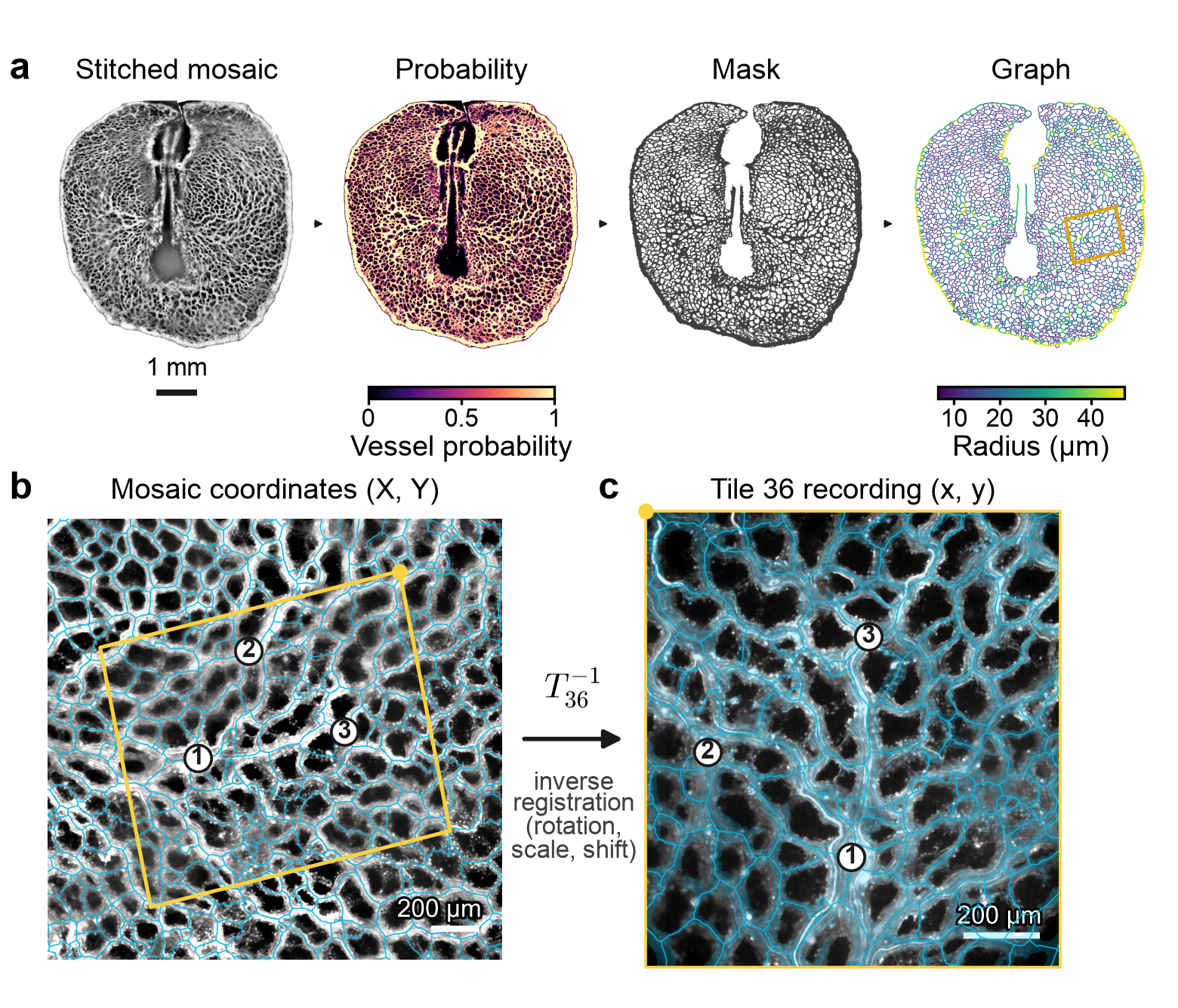}
    \caption{\textbf{Vascular segmentation and network extraction.}
    \textbf{a,} Stitched mosaic, vessel-probability map obtained using \textit{ilastik}, binary vascular mask, and vectorized graph colored by vessel radius.
    \textbf{b,} Portion of the reconstructed graph in mosaic coordinates $(X,Y)$. The yellow outline marks the footprint of tile 36.
    \textbf{c,} Vessel centerlines mapped into the original recording coordinates $(x,y)$ using the inverse registration transformation $T_{36}^{-1}$. Numbered markers identify corresponding locations in (b) and (c).}
    \label{fig:segmentation}
\end{figure}

We segmented the mosaic using \textit{ilastik} and vectorized the resulting vascular mask to obtain a spatial graph (Fig.~\ref{fig:segmentation}a). Graph nodes correspond to vessel junctions and termini, while edges retain the centerline geometry, length, and radius of each vessel segment. Regions in which individual vessels or their connectivity could not be resolved were excluded. The reconstructed network comprised 9,016 vessel segments connecting 5,904 nodes and spanned approximately $5.6\times6.0$\,mm. The median segment length and radius were 55 and 16\,$\mu$m, respectively.

Although the vascular graph was extracted from the stitched mosaic, flow measurements were obtained from the original recordings. For each tile, we applied the inverse registration transformation to recover the vessel centerlines in the recording's original coordinate system (Fig.~\ref{fig:segmentation}b,c). Across the network, 92\% of vessel segments were imaged in at least one recording and 71\% in multiple recordings.

\subsection*{Vessel-resolved measurements of mean and pulsatile blood flow}

\begin{figure}[!htbp]
\centering
\includegraphics[width=\linewidth]{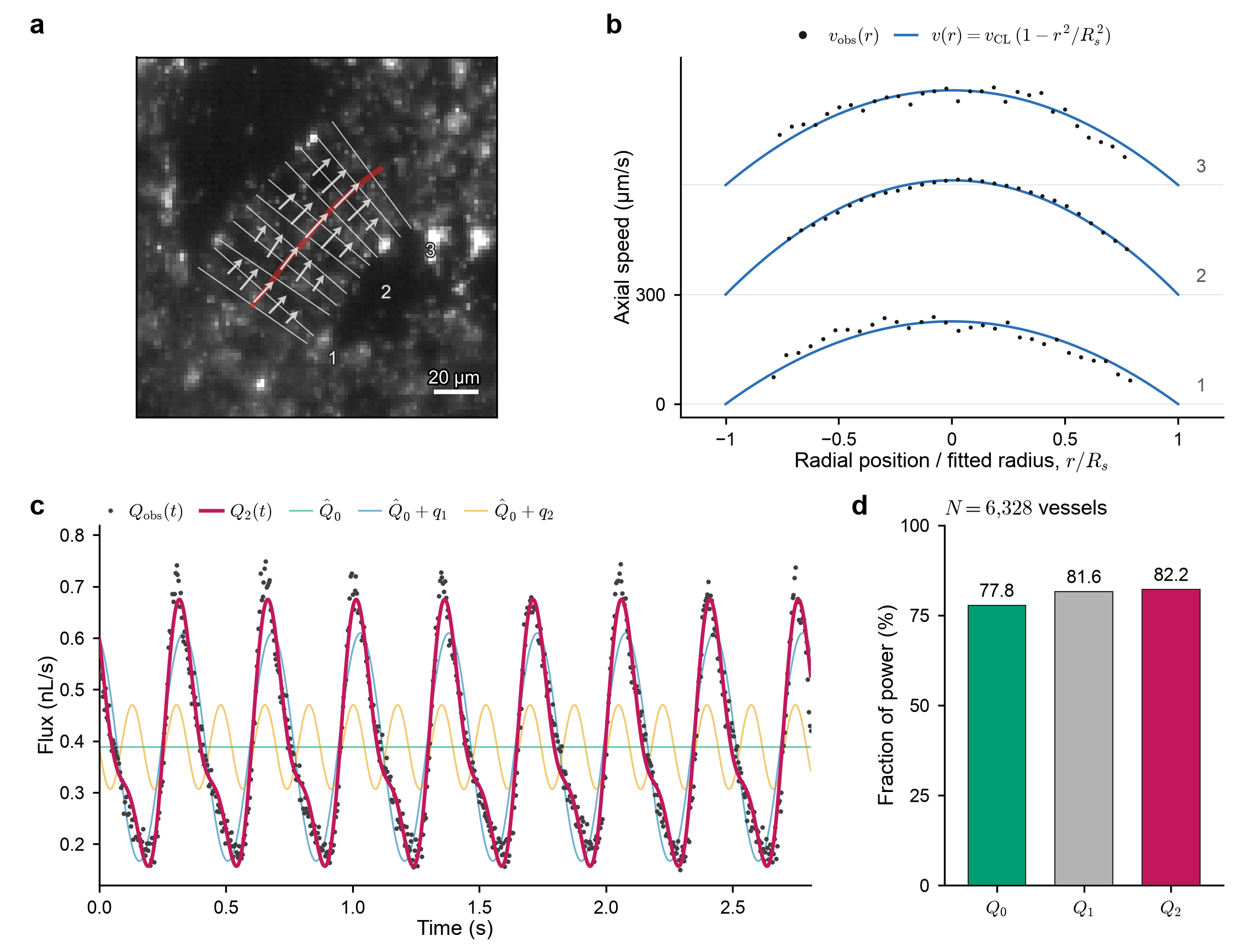}
\caption{\textbf{Estimation of vessel-resolved mean and pulsatile blood flow.}
\textbf{a,} Representative vessel showing its centerline (red), transverse sampling locations, and optical-flow velocity vectors.
Numbered locations correspond to the velocity profiles in (b).
\textbf{b,} Time-averaged axial velocity profiles at three cross-sections, with independently fitted parabolic profiles (blue).
The transverse coordinate $r$ is measured from the vessel centerline and normalized by the fitted effective radius $R_s$ of each cross-section.
Profiles are vertically offset for clarity.
The vessel-level mean flux, $\bar Q=0.39$\,nL/s, was estimated from 54 cross-sectional measurements.
\textbf{c,} Frame-resolved volumetric flux $Q_{\mathrm{obs}}(t)$ (black points), estimated from axial velocity profiles with radii fixed at their time-averaged values.
The fitted waveform $Q_2(t)$ (magenta) comprises the mean and the first two cardiac harmonics.
The individual harmonic contributions are shown added to the mean, $\hat Q_0+q_1$ and $\hat Q_0+q_2$.
\textbf{d,} Fraction of total signal power captured by fits containing the mean alone ($Q_0$), the mean and fundamental ($Q_1$), and the mean and first two harmonics ($Q_2$), pooled across 6,328 vessels measured in tiles 7--53.}
\label{fig:flow_estimation}
\end{figure}

The reconstructed vessel geometry allowed us to estimate blood flow directly from the original high-speed recordings (Fig.~\ref{fig:flow_estimation}a).
Because the measured particle velocities represent two-dimensional projections of flow through three-dimensional vessels, volumetric flux cannot be determined directly from the observed velocities.
We therefore estimated mean flux from transverse velocity profiles, which were approximately parabolic in well-resolved vessel segments (Fig.~\ref{fig:flow_estimation}b).
In the example shown, measurements across 54 cross-sections yielded a vessel-level mean flux of $0.39\,\mathrm{nL/s}$.

The same recordings also resolved temporal variations in flow over successive cardiac cycles.
We combined the time-dependent velocity measurements with the mean flux estimate to obtain a flow waveform for each vessel.
Figure~\ref{fig:flow_estimation}c shows one such waveform and its decomposition into the first two cardiac harmonics.
The second harmonic captures features of the waveform that are absent from a sinusoidal representation of flow.
Across the 6,328 measured vessels, the mean component alone accounted for $77.8\%$ of the total signal power, increasing to $81.6\%$ with the fundamental and $82.2\%$ with the first two harmonics (Fig.~\ref{fig:flow_estimation}d).

\subsection*{Pulsatile flow varies along connected vascular paths}

\begin{figure}[!htbp]
    \centering
    \includegraphics[width=\linewidth]{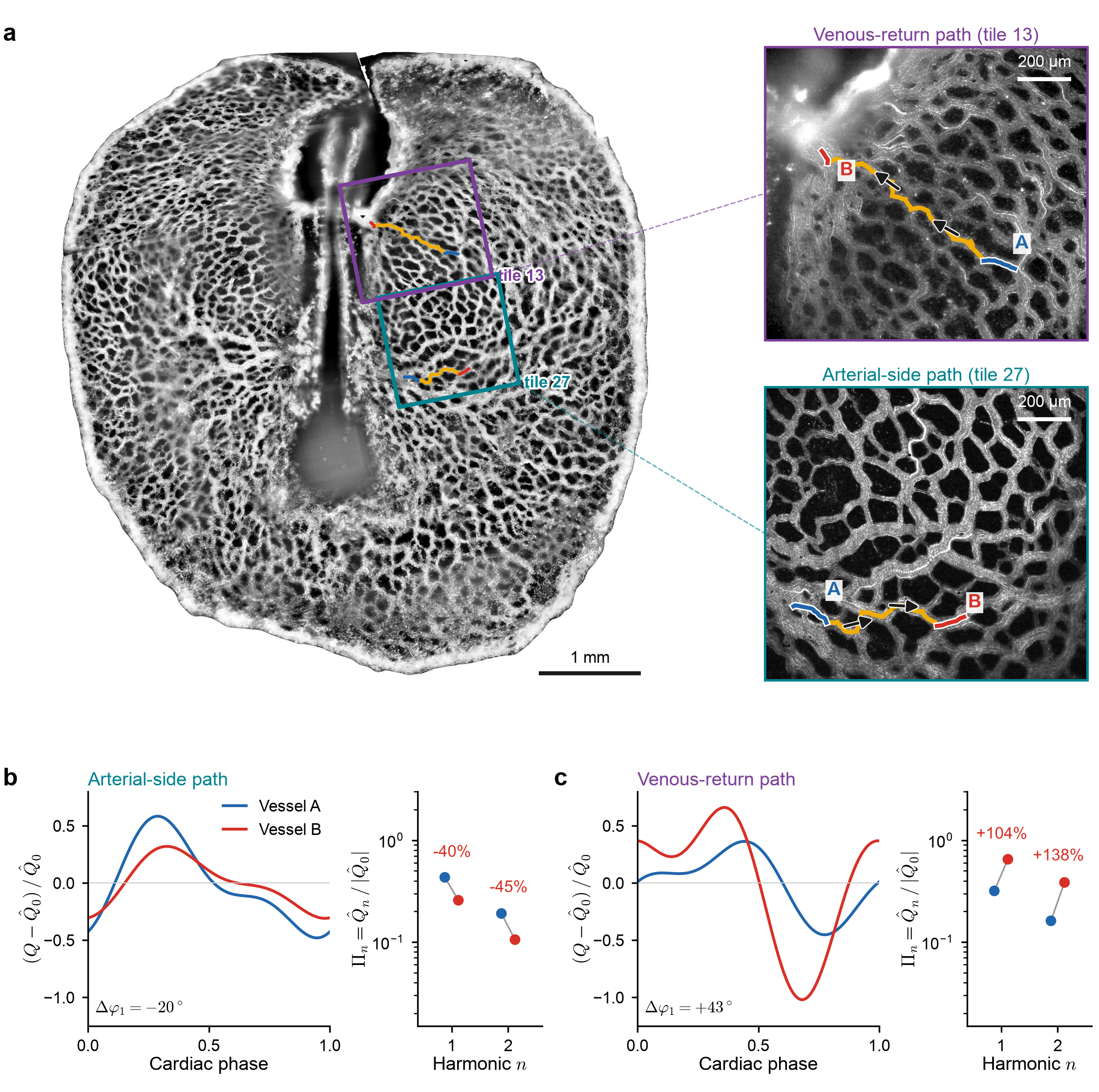}
    \caption{\textbf{Local variation in pulsatile flow along connected vascular paths.}
    \textbf{a,} Locations of the two vascular paths within the yolk sac. Vessels A (blue) and B (red) are connected by the path shown in yellow. Arrows indicate the direction of mean flow.
    \textbf{b,c,} Flow waveforms and relative amplitudes of the first two cardiac harmonics along the arterial-side (b) and venous-return (c) paths.. The waveforms are normalized by their respective mean fluxes and centered at zero. Percentages indicate the change in relative harmonic amplitude from A to B. The phase difference $\Delta\varphi_1$ is positive when the fundamental oscillation in B leads that in A.}
    \label{fig:local_pulsatility}
\end{figure}

Flow waveforms varied even between vessels connected by short vascular paths. Figure~\ref{fig:local_pulsatility} shows two such paths, one on the arterial side of the plexus and the other along the venous return. In each case, the vessels were recorded simultaneously, allowing direct comparison of their harmonic amplitudes and relative phases.

Along the arterial-side path, the amplitudes of the first two harmonics relative to mean flow decreased by 40\% and 45\%, respectively.
The fundamental oscillation in vessel B lagged that in vessel A by $20^\circ$ (Fig.~\ref{fig:local_pulsatility}b).
Along the venous-return path, we observed the opposite behavior: both relative harmonic amplitudes more than doubled, and the fundamental oscillation in B led that in A by $43^\circ$ (Fig.~\ref{fig:local_pulsatility}c).
Thus, even as blood flows toward the vitelline veins, its relative pulsatility can increase. The accompanying phase advance suggests that the venous boundary is itself a source of pulsatile forcing within the plexus.

\subsection*{Computational domain and vascular boundaries}

\begin{figure}[!htbp]
    \centering
    \includegraphics[width=\linewidth]{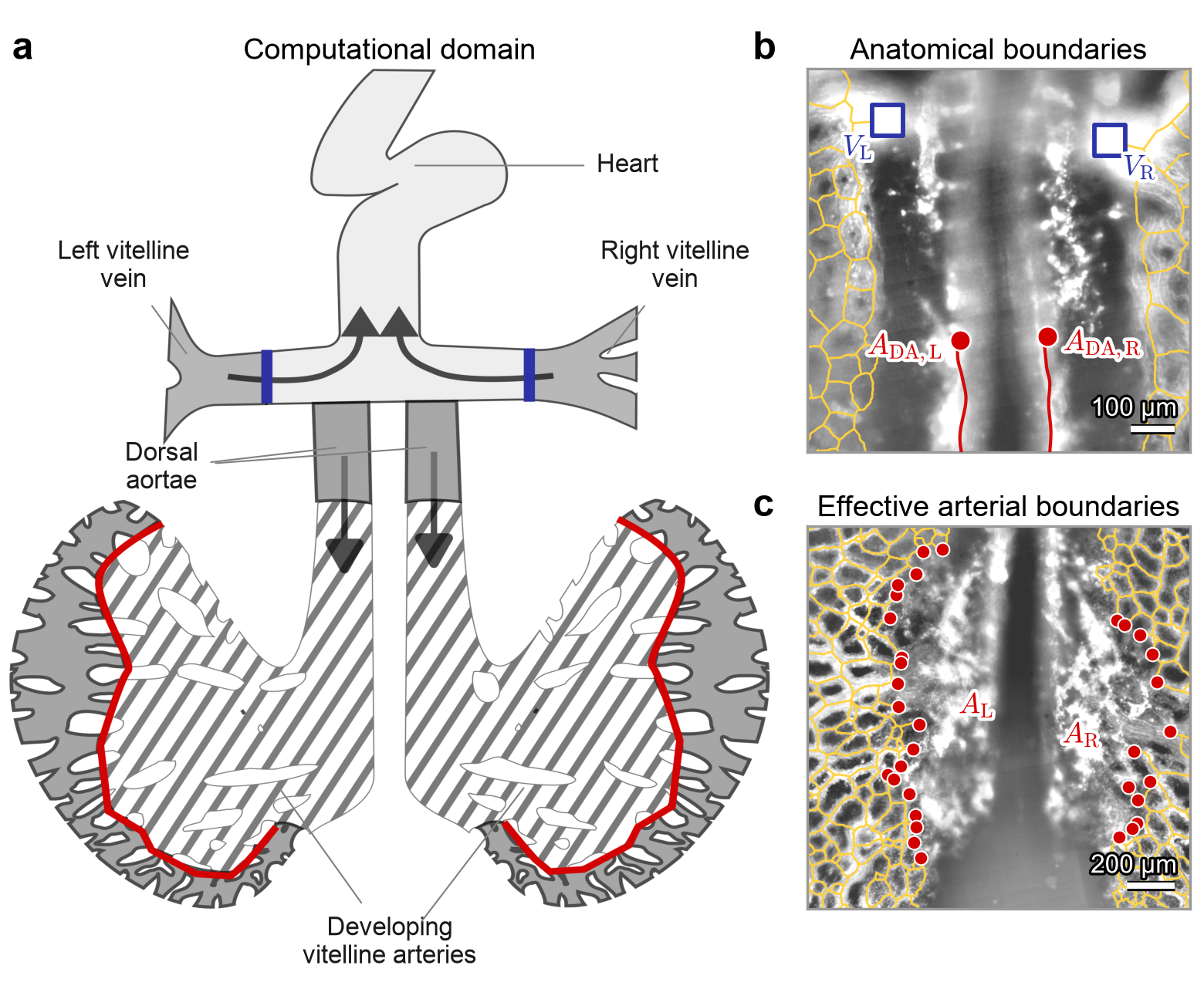}
    \caption{\textbf{Computational domain and vascular boundary identification.}
    \textbf{a,} Schematic of the embryonic yolk-sac circulation showing the dorsal aortae, developing vitelline arteries, and venous return to the heart.
    The computational domain is restricted to the resolved vascular plexus, excluding the central regions where individual arterial connections could not be reliably reconstructed.
    \textbf{b,} Anatomical arterial and venous vessels imaged simultaneously in tile 14.
    The two dorsal aortae ($A_{\mathrm{DA},L}$ and $A_{\mathrm{DA},R}$, red) and left and right vitelline veins ($V_L$ and $V_R$, blue) are indicated, together with the resolved vascular network (yellow).
    \textbf{c,} Effective arterial boundary nodes (red) defined along the interface between the resolved vascular plexus (yellow) and the unresolved central arterial region.
    Boundary nodes were grouped into left ($A_L$) and right ($A_R$) arterial regions.}
    \label{fig:network_boundaries}
\end{figure}

The reconstructed vascular network did not resolve all connections between the dorsal aortae and the surrounding yolk-sac plexus.
Near the dorsal aortae, densely packed developing vitelline arteries could not be reliably distinguished or assigned to individual junctions (Fig.~\ref{fig:network_boundaries}a).
Because flow depends on connectivity throughout the network, incorrectly assigning these connections could alter the inferred flow distribution far beyond the unresolved region.

The two dorsal aortae and the left and right vitelline veins were recorded simultaneously in tile 14, providing measurements of the major arterial and venous flows within a common temporal reference (Fig.~\ref{fig:network_boundaries}b).
However, the intervening connections between the dorsal aortae and the resolved plexus could not be traced reliably.
The anatomical arterial inlets therefore could not be connected unambiguously to the reconstructed network.

Rather than introduce uncertain arterial connections, we restricted the computational domain to the resolved plexus and defined effective arterial boundaries where individual vessels entered the unresolved central region (Fig.~\ref{fig:network_boundaries}c).
The resulting computational domain comprised 8,467 vessel segments.
The 33 arterial boundary nodes were grouped into left and right arterial regions, $A_L$ and $A_R$, while the left and right vitelline veins defined the venous boundaries, $V_L$ and $V_R$.
The effective arterial boundaries thus represent connections to unresolved upstream vasculature, rather than the anatomical arterial inlets themselves.
The fluxes through these boundaries were therefore not prescribed from the dorsal-aorta measurements, but inferred from the measured flow distribution throughout the resolved network.
\subsection*{Reconstruction of mean and pulsatile blood flow}

The overlap between image tiles provided repeated measurements of vessels throughout the yolk sac.
These measurements differed both in mean flux and in the timing of flow oscillations relative to the start of each recording.
We reconciled the mean and pulsatile components of flow separately.

We first combined the mean-flow measurements from all 53 tiles, comprising 13,711 recordings of 7,009 vessels.
Measurements of the same vessel were averaged using inverse-variance weights, as described in Materials and Methods.
Because the resulting mean fluxes did not necessarily satisfy conservation of flow at vascular junctions, we sought the flow distribution closest to these measurements that conserved flux at interior junctions.
We additionally imposed equal total arterial inflow into the left and right sides of the plexus.
The reconciled mean-flow field yielded a total inflow of $5.75\,\mathrm{nL/s}$.
This reconstruction did not assume any relationship between vessel geometry and hydraulic resistance.

\begin{figure}[!htbp]
    \centering
    \includegraphics[width=\linewidth]{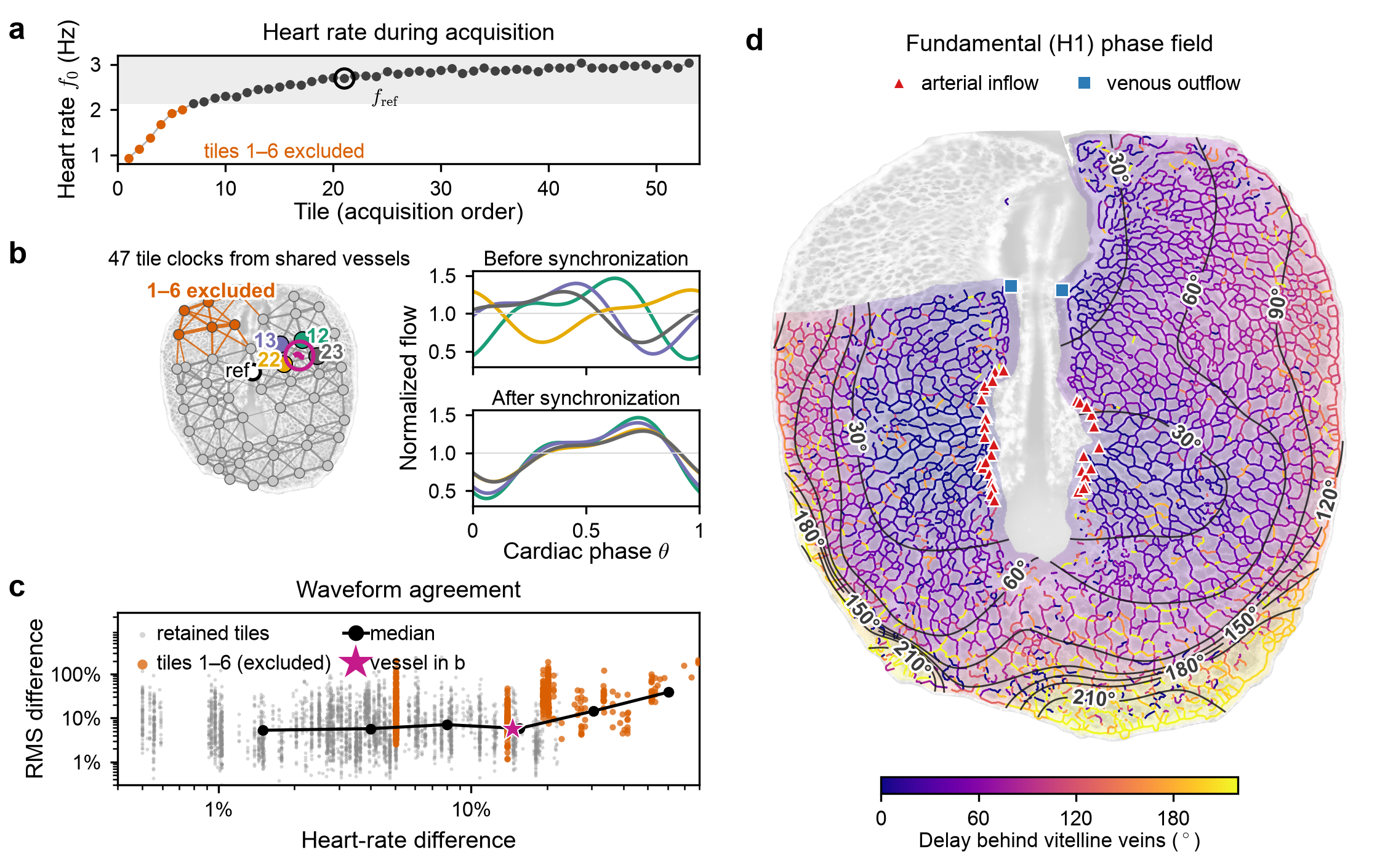}
    \caption{\textbf{Reconstruction of pulsatile flow across the vascular network.}
    \textbf{a,} Heart rate during image acquisition.
    The circled point denotes the reference tile, and the shaded region indicates the retained range of heart rates.
    Tiles 1--6 (orange) were excluded from the pulsatile reconstruction.
    \textbf{b,} Graph of overlapping tiles connected by shared vessels.
    The normalized flow waveforms of a vessel recorded in four tiles are shown before and after cardiac phase synchronization.
    \textbf{c,} Weighted root-mean-square difference between synchronized, normalized waveforms of repeatedly measured vessels as a function of the relative difference in heart rate between recordings.
    Orange points indicate vessels with measurements in at least one of the excluded tiles.
    Black markers denote binned medians, and the star identifies the vessel shown in (b).
    For this comparison, phase synchronization was performed using all 53 tiles.
    \textbf{d,} Spatial distribution of the first-harmonic phase lag relative to the vitelline veins.
    Contours represent spatial smoothing of the reconstructed phase field for visualization.
    Triangles and squares indicate arterial and venous boundary nodes, respectively.}
    \label{fig:flow_reconciliation}
\end{figure}

Combining the pulsatile measurements presented two additional challenges: the heart rate varied over the course of image acquisition, and each recording had an independent cardiac phase reference.
The measured heart rate increased from 0.93 to 3.03 Hz during acquisition (Fig.~\ref{fig:flow_reconciliation}a).
We selected tile 21 as the reference recording, with a heart rate of $f_{21}=2.69\,\mathrm{Hz}$, and restricted the pulsatile reconstruction to the 47 tiles whose heart rates were within $21\%$ of this frequency.
Measurements from the first six tiles were therefore excluded from the pulsatile reconstruction but retained in the mean-flow reconstruction.

We used repeated measurements of individual vessels to establish a common cardiac phase across the retained tiles.
We jointly determined the phase offsets of the tiles by minimizing weighted differences in first-harmonic phase between repeated measurements.
The phase offset of each tile was applied to both harmonics of every vessel within that tile.

Figure~\ref{fig:flow_reconciliation}b shows the normalized flow waveforms of a vessel recorded in four different tiles before and after synchronization.
Although the recordings were acquired at different heart rates, their waveforms agreed closely after alignment.
Across repeatedly measured vessels, the median RMS difference between normalized waveforms remained below $10\%$ for heart-rate differences of less than $10\%$ and increased at larger differences in heart rate (Fig.~\ref{fig:flow_reconciliation}c).

After synchronization, we averaged the harmonic coefficients of each vessel using the inverse-variance weights of the corresponding recordings.
We then rescaled these coefficients by the ratio of the reconciled mean flux to the combined mean flux across all eligible recordings.
Pulsatile flow was not reconstructed in 787 vessels for which this rescaling would reverse the direction of mean flow, or in 37 vessels for which the ratio could not be determined.
The resulting reconstruction comprised flow waveforms in 5,214 vessels, represented over a common cardiac cycle at the reference heart rate.

The synchronized waveforms revealed a spatial variation in the timing of pulsatile flow across the yolk sac (Fig.~\ref{fig:flow_reconciliation}d).
On both sides of the embryo, the first-harmonic phase lag relative to the vitelline veins generally increased toward the periphery of the plexus, exceeding $150^\circ$ along parts of the vascular margin.

\subsection*{Modeling pulsatile flow in a compliant vascular network}

We next investigated whether the reconstructed distribution of mean flow and the spatial variation in cardiac pulsatility could be explained by vascular geometry and vessel-wall elasticity.

\begin{figure}[!htbp]
    \centering
    \includegraphics[width=\linewidth]{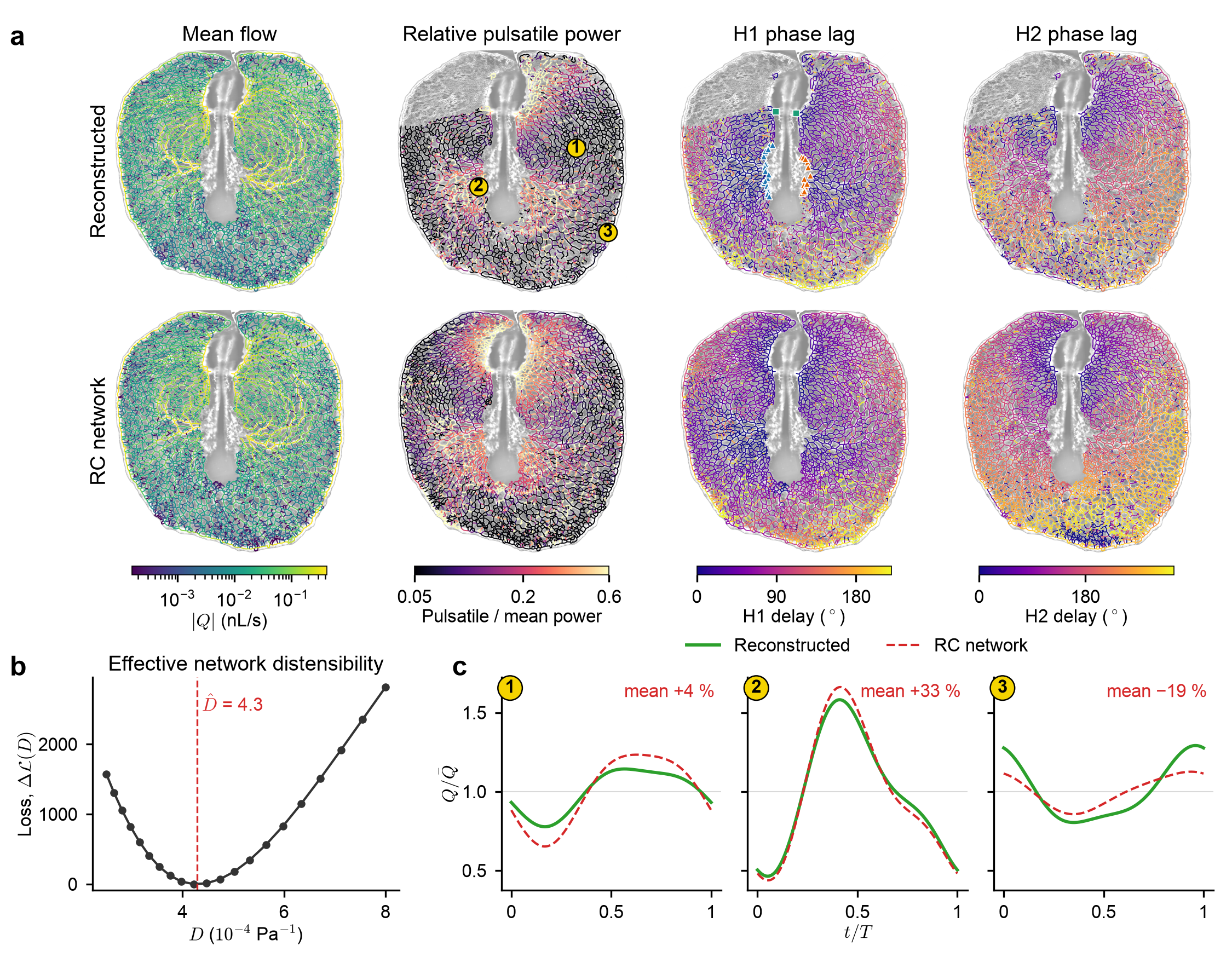}
    \caption{\textbf{Mean and pulsatile flow in a compliant vascular network.}
    \textbf{a,} Comparison of the reconstructed flow field (top) and the distributed-RC network model (bottom).
    From left to right: magnitude of mean flux, ratio of pulsatile to mean power, and phase lags of the first and second cardiac harmonics relative to the vitelline veins.
    Pulsatile power is defined as the mean-square oscillatory flux.
    Blue and orange triangles indicate the left and right arterial boundary nodes, respectively, and green squares indicate the venous boundaries.
    \textbf{b,} Model loss relative to its minimum as a function of the uniform vessel distensibility.
    The dashed line indicates the minimum at $D\approx4.3\times10^{-4}\,\mathrm{Pa}^{-1}$.
    \textbf{c,} Reconstructed (green) and simulated (dashed red) flow waveforms at the three vessels indicated in (a), each normalized by its respective mean flux.
    Percentages indicate the relative difference between the simulated and reconciled mean fluxes.}
    \label{fig:rc_model}
\end{figure}

We first represented the vascular network as a collection of passive hydraulic resistances, with the resistance of each vessel determined by its measured radius and length.
Prescribing the boundary fluxes obtained from the reconciled mean-flow field, we calculated the distribution of mean flow throughout the network.
The predicted fluxes within the plexus were thus determined by vascular geometry and the imposed boundary fluxes, without fitting the mean flow of individual vessels.
The model recovered several of the high-flux pathways within the plexus and along the vascular margin (Fig.~\ref{fig:rc_model}a).
Across the 7,009 vessels with flow measurements, the predicted and reconciled fluxes had a Pearson correlation of 0.93 and a Spearman rank correlation of 0.78 in their magnitudes, with agreement in mean-flow direction in 85\% of vessels.

To model the pulsatile component of flow, we extended the passive network to include vessel-wall compliance.
Each vessel was represented as a distributed resistance--compliance (RC) element, allowing blood to be stored and released along its length over the cardiac cycle.
We characterized vessel-wall elasticity by the areal distensibility,
\begin{align}
    D=\frac{1}{A}\frac{\partial A}{\partial P_{\mathrm{tm}}},
\end{align}
where $A$ is the vessel cross-sectional area and $P_{\mathrm{tm}}$ is the transmural pressure.
We assigned all vessels the same distensibility, so that differences in hydraulic resistance and compliance arose from differences in vessel geometry.

The pulsatile boundary conditions were obtained from the reconstructed flow field.
At each of the 33 arterial boundary nodes, we prescribed the reconciled mean inflow multiplied by a normalized waveform estimated from nearby vessels.
At the two venous boundaries, $V_L$ and $V_R$, we multiplied the respective mean outflows by a common normalized waveform obtained by averaging the measurements in the two vitelline veins.
The boundary fluxes and waveforms were held fixed, leaving $D$ as the only adjustable physical parameter of the model.

We evaluated the model over a range of distensibilities and compared the resulting pulsatile flow with the reconstructed waveforms in 4,968 vessels.
To separate differences in pulsatility from differences in mean flux, we compared waveforms normalized by their respective mean fluxes.
The loss accounted for differences in the phases of the first two cardiac harmonics and the relative amplitude of the first harmonic, with additional variance parameters fitted for each of the three contributions.
The loss was minimized at $D\approx4.3\times10^{-4}\,\mathrm{Pa}^{-1}$ (Fig.~\ref{fig:rc_model}b).

Despite the assumption of uniform distensibility, the model reproduced several spatial features of the reconstructed flow field (Fig.~\ref{fig:rc_model}a).
Both fields exhibited regions of enhanced and reduced pulsatility, as well as increasing phase lags toward the vascular margin.
The greater spatial variation in phase of the second harmonic was also apparent in the model.

Agreement in mean flux and agreement in the pulsatile waveform varied among individual vessels.
For the three vessels shown in Fig.~\ref{fig:rc_model}c, the simulated mean flux differed from the reconciled mean flux by $+4\%$, $+33\%$, and $-19\%$, respectively.
Despite the difference in mean flux, the normalized waveform in vessel 2 was well reproduced by the model.
Larger discrepancies in pulsatile amplitude and waveform shape were apparent in vessels 1 and 3.
These comparisons show that a network model incorporating vessel geometry and uniform elasticity can account for broad spatial features of cardiac pulsatility, although local flow waveforms are not fully captured.

\section*{Materials and Methods}

\subsection*{Embryo preparation and high-speed imaging}

We analyzed previously acquired high-speed fluorescence microscopy recordings of the yolk-sac vasculature of a Japanese quail (\textit{Coturnix japonica}) embryo at the 21-somite stage (approximately Hamburger--Hamilton stage 14).
Blood flow was visualized using circulating fluorescent tracer microspheres.

High-speed imaging was performed using a Photron FASTCAM-APX RS 250K camera and a 10$\times$ objective.
Images were acquired at 250 frames\,s$^{-1}$, corresponding to a frame interval of 4\,ms.

Because each field of view covered only a portion of the yolk-sac vasculature, the network was imaged as a sequence of spatially overlapping recordings (Fig.~\ref{fig:acquisition}a).
Individual recordings typically contained 800 frames (3.2\,s), with some extending to 2000 frames (8\,s).
The complete acquisition comprised 53 recordings, totaling approximately 230\,s of high-speed imaging distributed over 3.7\,h.
Between recordings, the microscope stage was repositioned along a serpentine acquisition path (Fig.~\ref{fig:acquisition}c).

Over the course of acquisition, both vascular geometry and cardiac dynamics changed, complicating the reconstruction of geometry and blood flow across the entire yolk sac.

\subsection*{Image projection and spatial registration}
\label{sec:registration}

To obtain structural images of the local vasculature, each high-speed recording was reduced to a two-dimensional maximum-intensity projection, followed by a $3\times3$ median filter to suppress isolated pixel noise (Fig.~\ref{fig:acquisition}a). The resulting projections were registered into a common embryo-centered coordinate system to reconstruct the vascular mosaic.

Registration was complicated by tissue growth and vascular remodeling during the approximately 3.7\,h acquisition, which introduced differences in scale and local vascular geometry between overlapping recordings. Tiles were positioned interactively using a custom \texttt{napari}-based interface, with a common rotation applied to align the mosaic with the embryo's principal anatomical axes. Where translation alone was insufficient, affine corrections were estimated by least-squares fitting to manually identified corresponding vascular landmarks (Fig.~\ref{fig:acquisition}b). Such corrections were applied to 31 of the 53 tiles, with the largest adjustments occurring in the earliest recordings. Alignment was inspected and refined iteratively across overlapping regions.

Before mosaic assembly, the projections were flat-field corrected and intensity-normalized to reduce differences in illumination and contrast between recordings. The transformed images were combined using distance-weighted feathering in overlapping regions, producing a continuous structural representation of the imaged yolk-sac vasculature (Fig.~\ref{fig:acquisition}c).
\subsection*{Vessel segmentation and network extraction}
\label{sec:segmentation}

The registered structural mosaic was segmented using a two-stage pixel classifier in \textit{ilastik} \citep{berg2019ilastik} (Fig.~\ref{fig:segmentation}a). The resulting vessel-probability map was converted to a binary mask using hysteresis thresholding, followed by morphological processing and manual correction to remove spurious structures, recover missed vessels, and preserve network connectivity. Regions within the embryo proper that could not be reliably segmented were excluded.

The corrected mask was vectorized using \texttt{net3}, a Python~3 implementation and extension of the NET framework \citep{lasser_net_2017}. The resulting spatial graph represents vessel junctions and termini as vertices and individual vessel segments as edges, retaining their centerlines, lengths, and local radius profiles obtained from the segmented geometry. Short spurious branches were pruned, and the largest connected component was retained. The graph was subsequently inspected and corrected interactively to resolve segmentation and connectivity errors.

Although segmentation and vectorization were performed in the common mosaic coordinate system, blood-flow measurements were extracted from the original high-speed recordings. For each image tile, the corresponding portion of the vascular graph was transformed back into its original image coordinates using the inverse affine registration transformation (Fig.~\ref{fig:segmentation}b,c). This provided vessel centerlines and local geometry in the coordinate system of each recording, enabling direct sampling of the time-resolved image data without resampling the original videos. 

\subsection*{Vessel-resolved estimation of blood flow}

We computed a dense optical-flow field between consecutive frames of each recording using Farnebäck's method.
Pixel displacements were converted to velocities using the spatial calibration of $1.7\,\mu\mathrm{m}$ per pixel and the frame interval of 4\,ms.

For each vessel visible in a tile, we sampled the measured velocity field at cross-sections perpendicular to the vessel centerline.
At each cross-section, velocities were projected onto the local centerline tangent, with their signs defined by a fixed orientation of the vessel.
We fitted the time-averaged axial velocity profile at each cross-section $s$ with a parabola,
\begin{align}
    v_s(r) = v_{\mathrm{CL},s}\left(1-\frac{r^2}{R_s^2}\right),
\end{align}
where $r$ is the transverse coordinate measured from the centerline, and the centerline velocity $v_{\mathrm{CL},s}$ and effective radius $R_s$ were both fitted.

A cross-section was retained if the coefficient of determination of the fit was at least 0.5, the fitted radius was between 0.5 and 2 times the radius obtained from segmentation, and the mean flow direction agreed with that of the majority of cross-sections within the vessel.
We required at least two retained cross-sections to estimate flow in a vessel.

Assuming an axisymmetric velocity profile, we integrated the fitted profile over each cross-section and averaged the resulting fluxes to estimate the mean volumetric flux,
\begin{align}
    \overline Q_{e,\alpha}
    =
    \frac{\pi}{2}
    \left\langle R_s^2 v_{\mathrm{CL},s}\right\rangle_s,
\end{align}
where $e$ denotes the vessel, $\alpha$ denotes the tile, and $\langle\cdot\rangle_s$ denotes an average over the retained cross-sections.

To resolve temporal variations in flow, we fitted the axial velocity profiles at each frame while holding the radii fixed at their time-averaged estimates.
The instantaneous volumetric flux was then estimated as
\begin{align}
    Q^{\mathrm{obs}}_{e,\alpha}(t_j)
    =
    \frac{\pi}{2}
    \left\langle R_s^2 U_s(t_j)\right\rangle_s,
\end{align}
where $U_s(t_j)$ is the fitted centerline velocity at cross-section $s$ and time $t_j$.
Holding the radii fixed prevented fluctuations in the fitted radius arising from measurement noise from entering the flux estimates.

We estimated the cardiac frequency $f_\alpha$ of each tile by identifying a common spectral peak in the periodograms of the flux time series of its 12 longest measured vessels.

\subsection*{Measurement reconciliation}

A tile, $\alpha$, measures the flow rate on a given edge.
Raw timeseries consistently contain a mean, and the first two multiples of the heartbeat frequency $f_\alpha$ (figure~\ref{fig:flow_estimation}).
We fit the raw data with
\begin{align}
    Q^{\mathrm{fit}}_{e,\alpha}(t)
    =
    \overline Q^{\mathrm{fit}}_{e,\alpha}
    +
    \sum_{n=1}^{2}
    \left[
        \hat Q^{\mathrm{fit}}_{e,\alpha,n}e^{in\omega_\alpha t}
        + \mathrm{c.c.}
    \right],
\end{align}
where $\omega_\alpha=2\pi f_\alpha$ and $\overline Q_{e,\alpha}^{\mathrm{fit}}=\hat Q_{e,\alpha,0}^{\mathrm{fit}}$.

We estimate each recording's uncertainty from the difference between the observed and fitted flow, before clock alignment.
For recording $(e,\alpha)$, define the residual
\begin{align}
    r_{e,\alpha}(t_j)
    =
    Q^{\mathrm{obs}}_{e,\alpha}(t_j)
    -
    Q^{\mathrm{fit}}_{e,\alpha}(t_j).
\end{align}
With $M_{e,\alpha}$ observed samples in the timeseries, the estimated residual variance is
\begin{align}
    s_{e,\alpha}^2
    =
    \frac{1}{M_{e,\alpha}-5}
    \sum_{j=1}^{M_{e,\alpha}} r_{e,\alpha}(t_j)^2.
\end{align}
The five fitted parameters are the mean and two complex pairs, with the tile frequency held fixed.
Assuming independent measurement errors, we estimate the variance of the fitted mean as
\begin{align}
    \sigma_{e,\alpha}^2
    \approx
    \frac{s_{e,\alpha}^2}{M_{e,\alpha}}.
\end{align}
We use this uncertainty to weight the recordings during both clock alignment and waveform averaging.

We align the tile recordings to a common cardiac clock set by tile 21, which contains a strong heartbeat reference signal measured along the dorsal aortae.
Its angular frequency is $\omega_{21}=2\pi f_{21}$.
Each recording is fitted at its own frequency $f_\alpha$, and the synchronized cycles are then represented at $\omega_{21}$.

We synchronize the tile clocks by comparing measurements of the same vessel.
As in figure~\ref{fig:flow_reconciliation}, tiles 7 and above have comparable measured heartbeat frequencies, within $21\%$ of $f_{21}$.
We omit tiles 6 and under (the top left region of the plexus) from the phase alignment. 
We assume that differences in the first-harmonic phase of the same vessel across recordings arise from tile-dependent clock offsets and measurement error.
The offset $\psi_\alpha$ is shared by all vessels measured in tile $\alpha$.
Under the same noise assumptions, a small-error approximation gives the first-harmonic phase variance
\begin{align}
    \tau_{e,\alpha}^2
    \approx
    \frac{\sigma_{e,\alpha}^2}{2\left|\hat Q^{\mathrm{fit}}_{e,\alpha,1}\right|^2}.
\end{align}
A recording with a more clearly resolved first harmonic provides a more precise clock estimate.
For independent recordings, the variances of their phase errors add, so we weight each comparison by
\begin{align}
    w_{e,\alpha\beta}
    =
    \frac{1}{\tau_{e,\alpha}^2+\tau_{e,\beta}^2}.
\end{align}
Taking tile 21 as the reference, we estimate the offsets by minimizing
\begin{align}
    \underset{\{\psi_\alpha\}:\,\psi_{21}=0}{\operatorname{min}}
    \sum_e \sum_{\substack{\alpha,\beta\in\mathcal T_e\\ \alpha<\beta}} w_{e,\alpha\beta}
    \left[
        1-\cos\left((\arg\hat Q^{\mathrm{fit}}_{e,\alpha,1}-\psi_\alpha)-(\arg\hat Q^{\mathrm{fit}}_{e,\beta,1}-\psi_\beta)\right)
    \right],
\end{align}
where $\mathcal T_e$ contains the tiles retained for synchronized pulsatility that measure edge $e$.

The resulting offset is applied to every vessel within its tile, transforming the harmonic coefficients as
\begin{align}
    \hat Q^{\mathrm{fit}}_{e,\alpha,n}
    \longmapsto
    \hat Q^{\mathrm{fit}}_{e,\alpha,n}e^{-in\psi_\alpha}.
\end{align}
This preserves the waveform shape and phase differences between vessels within each tile, while leaving the mean unchanged.
We then represent all synchronized cycles at the reference frequency $\omega_{21}$.

We now combine the tile measurements of each vessel by a weighted average.
We assign each recording the weight $1/\sigma_{e,\alpha}^2$.
Using the phase-corrected coefficients, the combined estimates are
\begin{align}
    \hat Q^{\mathrm{av}}_{e,n}
    =
    \frac{
        \displaystyle\sum_{\alpha\in\mathcal T_{e,n}}
        \hat Q^{\mathrm{fit}}_{e,\alpha,n}/\sigma_{e,\alpha}^2
    }{
        \displaystyle\sum_{\alpha\in\mathcal T_{e,n}}
        1/\sigma_{e,\alpha}^2
    }.
\end{align}
Here $\mathcal T_{e,0}$ includes all eligible mean-flow measurements, including tiles 1--6, whereas $\mathcal T_{e,1}=\mathcal T_{e,2}=\mathcal T_e$ contains the recordings retained for synchronized pulsatility.
The weights are normalized over the contributing recordings for each coefficient.
We write $\overline Q_e^{\mathrm{av}}=\hat Q^{\mathrm{av}}_{e,0}$.

Finally, we project the combined mean flows onto the constraints of flow conservation at internal junctions and equal total left and right arterial inflows.
Let $\mathbf C$ contain the signed junction-conservation rows and the additional left--right symmetry row, so that these constraints are $\mathbf C\overline{\boldsymbol Q}^{*}=0$.

We approximate the covariance of the combined mean flows by a diagonal matrix $\boldsymbol\Sigma$.
Assuming independent errors between tile recordings of each vessel, its diagonal entries are
\begin{align}
    \boldsymbol\Sigma_{ee}
    =
    \left(
        \sum_{\alpha\in\mathcal T_{e,0}}
        \frac{1}{\sigma_{e,\alpha}^2}
    \right)^{-1}.
\end{align}
The uncertainty-weighted projection is
\begin{align}
    \overline{\boldsymbol Q}^{*}
    =
    \overline{\boldsymbol Q}^{\mathrm{av}}
    -
    \boldsymbol\Sigma\mathbf C^{\mathsf T}
    \left(
        \mathbf C\boldsymbol\Sigma\mathbf C^{\mathsf T}
    \right)^{+}
    \mathbf C\overline{\boldsymbol Q}^{\mathrm{av}},
\end{align}
where $+$ denotes the Moore--Penrose pseudoinverse.
These constraints act on the mean flow only.
We then rescale each combined waveform by the ratio of its reconciled mean to its combined mean:
\begin{align}
    Q_e^{*}(t)
    =
    \overline Q_e^{*} + \frac{\overline Q_e^{*}}{\overline Q_e^{\mathrm{av}}}\sum_{n=1}^{2}
    \left[
        \hat Q^{\mathrm{av}}_{e,n}e^{in\omega_{21}t}
        + \mathrm{c.c.}
    \right].
\end{align}
A positive rescaling preserves the harmonic phases and relative waveform shape while imposing the reconciled mean.
We retain the reconciled mean but do not reconstruct pulsatility where synchronized harmonics are unavailable or the rescaling ratio is negative or undefined.

\subsection*{Passive network model of mean flow}
\label{sec:passive_model}

To assess the contribution of vascular geometry to the reconstructed mean-flow distribution, we represent each vessel as a passive hydraulic resistance.
Assuming steady, laminar flow through cylindrical vessels, the conductance of vessel $e$ is given by

\begin{equation}
    g_e = \frac{\pi r_e^4}{8\mu L_e},
\end{equation}

where $r_e$ and $L_e$ are the image-derived vessel radius and length, respectively, and $\mu$ is the effective dynamic viscosity.

We approximate blood as a Newtonian fluid with spatially uniform effective viscosity.
There is relatively low hematocrit during early avian development, and the vessel radii are generally larger than standard capillary dimensions.
Previous measurements of embryonic blood rheology support an approximately Newtonian description under appropriate physiological conditions in larger embryonic vessels \citep{al2011rheology}.

We model the steady network pressure field by combining the conductance-weighted pressure--flow relations with conservation of mass at interior junctions.
The predicted flux in vessel $e$, oriented from node $u$ to node $v$, is
\begin{equation}
    Q_{e} = g_e(P_u-P_v).
\end{equation}

Boundary fluxes are prescribed from the conservation-consistent mean-flow field $\overline{\boldsymbol Q}^{*}$, without fitting individual vessel flows.
Since all vessels share the same effective viscosity, a uniform rescaling of $\mu$ rescales the pressure differences but leaves the predicted flow distribution unchanged.

\subsection*{Distributed-RC model of pulsatile flow}
\label{sec:rc_model}

To investigate how vascular compliance influences the spatial organization of cardiac pulsatility, we extend the passive resistance network to include distributed vessel-wall compliance.
In the low-inertia regime of the embryonic yolk-sac circulation, viscous resistance dominates fluid inertance.
We therefore represent each vessel as a distributed resistance--compliance (RC) element, allowing periodic storage and release of blood along its length.

The pressure--area compliance per unit length is defined as

\begin{equation}
    \chi_e =
    \frac{\partial A_e}
    {\partial P_{\mathrm{tm}}},
\end{equation}

where $A_e=\pi r_e^2$ is the vessel cross-sectional area, $P_{\mathrm{tm}}$ denotes transmural pressure, and the derivative is evaluated at the mean transmural pressure.

Letting $Q_e(z,t)$ and $P_e(z,t)$ denote volumetric flux and pressure perturbations, respectively, the linearized governing equations are

\begin{equation}
    \frac{\partial Q_e}{\partial z}
    +
    \chi_e
    \frac{\partial P_e}{\partial t}
    =0,
    \qquad
    \frac{\partial P_e}{\partial z}
    =-\frac{Q_e}{g_e L_e}.
    \label{eq:rc_governing}
\end{equation}

For harmonic components proportional to $e^{\mathrm{i}\Omega_n t}$, these equations give

\begin{equation}
    \frac{d^2\widehat P_e^{(n)}}{dz^2}
    =
    \gamma_{e,n}^2
    \widehat P_e^{(n)},
    \qquad
    \gamma_{e,n}^2
    =
    \frac{\mathrm{i}\Omega_n\chi_e}{g_e L_e}.
    \label{eq:rc_harmonic}
\end{equation}

We define $\xi_{e,n}=\gamma_{e,n}L_e$.
The exact distributed-RC solution gives the endpoint admittances

\begin{equation}
    Y_{e,n}^{\mathrm{s}}
    =
    g_e\xi_{e,n}\coth\xi_{e,n},
    \qquad
    Y_{e,n}^{\mathrm{t}}
    =
    g_e\xi_{e,n}\operatorname{csch}\xi_{e,n}.
    \label{eq:rc_admittance}
\end{equation}

These relate the complex fluxes entering the vessel at its endpoints to the corresponding nodal pressures.
Assembling the contributions of all vessels and enforcing flux conservation at interior junctions yields a complex linear system for each harmonic,

\begin{equation}
    \mathcal L^{(n)}
    \widehat{\mathbf P}^{(n)}
    =
    \widehat{\mathbf q}_{\mathrm{ext}}^{(n)},
    \label{eq:rc_network}
\end{equation}

where $\mathcal L^{(n)}$ is the assembled network admittance matrix and $\widehat{\mathbf q}_{\mathrm{ext}}^{(n)}$ contains the imposed external harmonic fluxes.
The resulting pressure field determines the simulated vessel-flow harmonics.
In the zero-frequency limit, the model reduces to the steady Poiseuille network.

Because vessel-wall mechanics are not directly measured, we assume a uniform effective areal distensibility,

\begin{equation}
    D =
    \frac{1}{A_e}
    \frac{\partial A_e}
    {\partial P_{\mathrm{tm}}},
    \qquad
    \chi_e=\pi r_e^2D.
    \label{eq:rc_distensibility}
\end{equation}

We additionally allow a complex effective distensibility $D_{\mathrm{eff}}=D(1-\mathrm{i}\eta)$ to represent dissipative wall response, with $\eta=0$ corresponding to purely elastic compliance.
The parameters $D$ and $\eta$ determine the frequency-dependent attenuation and phase response of the network.

For comparison with the composite harmonic field, all simulations are evaluated at the reference frequency $f_{\mathrm{ref}}=2.69\,\mathrm{Hz}$, with $\Omega_n=2\pi n f_{\mathrm{ref}}$.
This treats measurements acquired at different cardiac frequencies as the response of one effective network at the reference frequency, and the inferred parameters are conditional on that approximation.

\section*{Discussion}

We found that pulsatility can increase in the direction of blood flow, particularly along the venous return.
In the example in Fig.~\ref{fig:local_pulsatility}c, this increase was accompanied by a phase advance.
Across the plexus, the largest phase lags relative to the vitelline veins occurred near the vascular margin.
Much of this spatial variation was reproduced by the RC model, even though we assigned the same distensibility to every vessel.

At the 21-somite stage, the vitelline veins are still relatively short, and functional sinuatrial valves have yet to develop.
Atrial contraction can therefore drive transient retrograde flow into the veins \citep{manner2007early,ho2019organ}.
Pulsations in the venous return need not be remnants of the arterial pulse transmitted through the plexus.
The two pulses may instead interact within the plexus.
If the vessel radius changes little over a heartbeat, the temporal variation in flow also reflects that in wall shear stress.
The differences in waveform between vessels are of interest given the sensitivity of endothelial cells to the temporal characteristics of shear \citep{buschmann_pulsatile_2010,weijts_blood_2018}.

The capacitive function of the embryonic dorsal aorta has previously been shown to shape pulsatile flow in the developing circulation \citep{anton_pulse_2013}.
In models of adult microvascular networks, the extent of pulsatility damping depends on hydraulic resistance, vessel compliance, and cardiac frequency \citep{pan2022pulsatility}.
Here, we considered compliance within the developing plexus itself, rather than in the larger vessels supplying it.

The mechanical properties of vessels at this stage of development are difficult to measure directly.
Previous studies have fitted lumped-parameter models of the embryonic circulation to simultaneous measurements of pressure and flow in the dorsal aorta \citep{yoshigi1997characterization,lucitti2005arterial}.
Although we did not measure pressure, we could compare the flow waveforms predicted for different values of distensibility with those reconstructed throughout the plexus.
The best-fitting value of $4.3\times10^{-4}\,\mathrm{Pa}^{-1}$ is comparable in magnitude to the value of $3\times10^{-4}\,\mathrm{Pa}^{-1}$ used to model the embryonic zebrafish circulation \citep{anton_pulse_2013}.

In developing avian vessels, apparent blood viscosity depends on hematocrit, vessel caliber, and shear rate \citep{al2011rheology,ghaffari_simultaneous_2015}.
We used the same effective viscosity throughout the network and assumed a linear elastic response of the vessel walls.
For a given vascular geometry, an increase in viscosity can be compensated for by a corresponding decrease in distensibility without changing the predicted flow field.
The fitted value of $D$ is therefore conditional on the assumed viscosity.
Measurements of pressure and vessel-wall deformation would allow for a more direct characterization of vessel distensibility.

The accuracy of the passive network model also depends on how well the vessel geometry is resolved.
Under Poiseuille flow, the hydraulic conductance of a vessel scales with the fourth power of its radius.
Even small errors in the measured radii can therefore affect the predicted distribution of flow.
In contrast, the conservation constraint used to reconcile the mean-flow measurements requires knowledge of vessel connectivity but not of hydraulic resistance.

The 53 recordings span 3.7\,h, during which the embryo grows and its heart rate changes considerably.
Vessels measured at the beginning and end of acquisition are therefore not necessarily in the same physiological state.
Neither the conservation constraint nor the synchronization procedure corrects for changes in the underlying distribution of blood flow. The present analysis is restricted to a single embryo, and the variability in these flow patterns between embryos remains unknown.

We prescribed the boundary flows in our model without considering the response of the heart to changes in the vascular network.
As the plexus remodels, changes in its hydraulic resistance and compliance may affect cardiac function.
The development of cardiac valves may also substantially alter pulsatility in the venous return.
Studying these effects would require a model of the heart coupled to the vascular network.

An open question is whether differences in the amplitude and timing of flow pulsations between vessels precede differences in their remodeling.
Addressing this question would require tracking individual vessels and their flow waveforms over successive stages of development.

\section*{Data and code availability}
The data and code underlying this study are available from the corresponding author upon reasonable request.

\bibliographystyle{unsrtnat}
\bibliography{references}
\end{document}